\documentstyle[aps,prd,epsf,preprint]{revtex}
\tighten
\draft
\begin{document}
\preprint{\makebox{\begin{tabular}{r}
 							INFNFE-17-98 \\
 							BARI-TH/323-98 \\
							OUTP-98-94-P \\
							              \\
\end{tabular}}}
\title{Big bang nucleosynthesis limit on $N_\nu$}
\author{E. Lisi~$^a$\footnote{Eligio.Lisi@ba.infn.it}, 
        S. Sarkar~$^b$\footnote{S.Sarkar@physics.ox.ac.uk} and 
        F.L. Villante~$^c$\footnote{Villante@fe.infn.it}}
\address{$^a$Dipartimento di Fisica and Sezione INFN di Bari,
             Via Amendola 173, I-70126 Bari, Italy \\
         $^b$Theoretical Physics, University of Oxford, 
             1 Keble Road, Oxford OX1 3NP, UK\\
	$^c$Dipartimento di Fisica and Sezione INFN di Ferrara, 
             Via del Paradiso 12, I-44100 Ferrara, Italy }
\date{\today}
\maketitle
\bigskip
\begin{abstract}
Recently we presented a simple method for determining the correlated
uncertainties of the light element abundances expected from big bang
nucleosynthesis, which avoids the need for lengthy Monte Carlo
simulations. We now extend this approach to consider departures from
the Standard Model, in particular to constrain any new light degrees
of freedom present in the thermal plasma during nucleosynthesis. Since
the observational situation regarding the inferred primordial
abundances has not yet stabilized, we present illustrative bounds on
the equivalent number of neutrino species $N_\nu$ for various 
combinations of
individual abundance determinations. Our $95\%$ C.L. bounds on $N_\nu$
range between 2 and 4, and can easily be reevaluated using the
technique provided when the abundances are known more accurately.
\end{abstract}
\bigskip

%========================================================================
\section{Introduction}

The Standard Model (SM) contains only $N_\nu=3$ weakly interacting
massless neutrinos. However the recent experimental evidence for
neutrino oscillations \cite{nuoscrev} may require it to be extended to
include new superweakly interacting massless (or very light) particles
such as singlet neutrinos or Majorons. These do not couple to the
$Z^0$ vector boson and are therefore not constrained by the precision
studies of $Z^0$ decays which establish the number of $SU(2)_{\rm L}$
doublet neutrino species to be \cite{pdg98}
\begin{equation}
 N_{\nu} = 2.993 \pm 0.011.
\label{nnulep}
\end{equation}
However, as was emphasized some time ago \cite{history}, such
particles would boost the relativistic energy density, hence the
expansion rate, during big bang nucleosynthesis (BBN), thus increasing
the yield of $^4$He. This argument was quantified for new types of
neutrinos and new superweakly interacting particles \cite{chicago} in
terms of a bound on the {\em equivalent number of massless neutrinos}
present during nucleosynthesis:
\begin{equation}
 N_{\nu} = 3 + f_{\rm B, F} \sum_{i} \frac{g_{i}}{2} 
                                     \left(\frac{T_{i}}{T_{\nu}}\right)^4,
\label{Nnudef}
\end{equation}
where $g_i$ is the number of (interacting) helicity states, $f_{\rm
B}=8/7$ (bosons) and $f_{\rm F}=1$ (fermions), and the ratio
$T_{i}/T_{\nu}$ depends on the thermal history of the particle under
consideration \cite{oss81}. For example, $T_{i}/T_{\nu}\leq0.465$ for
a particle which decouples above the electroweak scale such as a
singlet Majoron or a sterile neutrino. However the situation may be
more complicated, e.g. if the sterile neutrino has large mixing with a
left-handed doublet species, it can be brought into equilibrium
through (matter-enhanced) oscillations in the early universe, making
$T_{i}/T_{\nu}\simeq1$ \cite{ekt92}. Moreover such oscillations can
generate an asymmetry between $\nu_{\rm e}$ and $\bar{\nu_{\rm e}}$,
thus directly affecting neutron-proton interconversions and the
resultant yield of $^4$He \cite{nuosc}. This can be quantified in
terms of the {\em effective} value of $N_\nu$ parametrizing the
expansion rate during BBN, which may well be below 3! Similarly,
non-trivial changes in $N_\nu$ can be induced by the decays
\cite{nudec} or annihilations \cite{nuann} of massive neutrinos (into
e.g. Majorons), so it is clear that it is a sensitive probe of new
physics.

The precise bound on $N_{\nu}$ from nucleosynthesis depends on the
adopted primordial elemental abundances as well as uncertainties in
the predicted values. Although the theoretical calculation of the
primordial $^4$He abundance is now believed to be accurate to within
$\pm0.4\%$ \cite{lt98}, its observationally inferred value as reported
by different groups \cite{he4old,he4new} differs by as much as
$\approx4\%$. Furthermore, a bound on $N_\nu$ can only be derived if
the nucleon-to-photon ratio $\eta\equiv\,n_{\rm N}/n_\gamma$ (or its
lower bound) is known, since the effect of a faster expansion rate can
be compensated for by the effect of a smaller nucleon density. This
involves comparison of the expected and observed abundances of other
elements such as D, $^3$He and $^7$Li which are much more poorly
determined, both observationally and theoretically. The most crucial
element in this context is deuterium which is supposedly always
destroyed and never created by astrophysical processes following the
big bang \cite{dhist}. Until relatively recently \cite{dhi,dlo}, its
primordial abundance could not be directly measured and only an
indirect upper limit could be derived based on models of galactic
chemical evolution. As reviewed in ref.\cite{bbnrev}, the implied
lower bound to $\eta$ was then used to set increasingly stringent
upper bounds on $N_\nu$ ranging from 4 downwards \cite{many},
culminating in one below 3 which precipitated the so-called ``crisis''
for standard BBN \cite{crisis}, and was interpreted as requiring new
physics.

However as cautioned before \cite{eens86}, there are large systematic
uncertainties in such constraints on $N_\nu$ which are sensitive to
our limited understanding of galactic chemical evolution. Moreover it
was emphasized \cite{kernan} that the procedure used earlier
\cite{many} to bound $N_\nu$ was statistically inconsistent since,
e.g., correlations between the different elemental abundances were not
taken into account. A Monte Carlo (MC) method was developed for
estimation of the correlated uncertainties in the abundances of the
synthesized elements \cite{monte,mc}, and incorporated into the
standard BBN computer code \cite{code}, thus permitting reliable
determination of the bound on $N_\nu$ from estimates of the primordial
elemental abundances. Using this method, it was shown \cite{nocrisis}
that the {\em conservative} observational limits on the primordial
abundances of D, $^4$He and $^7$Li allowed $N_\nu\leq4.53$ ($95\%$
C.L.), significantly less restrictive than earlier estimates. Similar
conclusions followed from studies using maximum likelihood (ML)
methods \cite{cst97,maxlik,hsbl97}.
However the use of the Monte Carlo method is computationally expensive
and moreover the calculations need to be repeated whenever any of the
input parameters --- either reaction rates or inferred primordial
abundances --- are updated. 

In a previous paper \cite{us} we presented
a simple method for estimation of the BBN abundance uncertainties and
their correlations, based on linear error propagation. To illustrate
its advantages over the MC+ML method, we used simple $\chi^2$
statistics to obtain the best-fit value of the nucleon-to-photon ratio
in the standard BBN model with $N_\nu=3$ and indicated the relative
importance of different nuclear reactions in determining the
synthesized abundances. In this work we extend this approach to consider 
departures from
$N_\nu=3$. We have checked that our results are consistent with
those obtained independently \cite{ot97} 
using the MC+ML method \cite{ot97,decay}
where comparison is possible. 

The essential advantage of our method is
that the correlated constraints on $N_\nu$ and $\eta$ can be easily
reevaluated using just a pocket calculator and the numerical tables 
provided, when the input nuclear reaction
cross-sections or inferred abundances are known better. We have in
fact embedded the calculations in a compact Fortran code, which is
available upon request from the authors, or from a
website \cite{website}. Thus observers
will be able to readily assess the impact of new elemental abundance
determinations on an important probe of physics beyond the standard
model.

%==========================================================================
\section{The Method}

In this section we recapitulate the basics of our method \cite{us} and
outline its extension to the case $N_\nu\neq3$.

\subsection{Basic Ingredients}

The method has both experimental and theoretical ingredients.  The
experimental ingredients are: (a) the inferred values of the
primordial abundances, $\overline{Y_i}\pm\overline\sigma_i$; and (b)
the nuclear reaction rates, $R_k\pm\Delta\,R_k$. We normalize
\cite{us} all the rates to a ``default'' set of values ($R_k\equiv1$),
namely, to the values compiled in Ref.~\cite{mc}, except for the
neutron decay rate, which is updated to its current value
\cite{pdg98}.

The theoretical ingredients are: (a) the calculated abundances $Y_i$;
and (b) the logarithmic derivatives
$\lambda_{ik}=\partial\ln{Y_i}/\partial\ln\,R_k$. Such functions have
to be calculated for generic values of $N_\nu$ and of
$x\equiv\log_{10}(\eta_{10})$, where $\eta_{10}=\eta/10^{-10}$. Note
that the fraction of the critical density in nucleons is given by
$\Omega_{\rm N}h^2\simeq\eta_{10}/273$, where $h\sim0.7\pm0.1$ is the
present Hubble parameter in units of $100$~km~s$^{-1}$~Mpc$^{-1}$, and
the present temperature of the relic radiation background is
$T_0=2.728\pm0.002$~K \cite{pdg98}.

The logarithmic derivatives $\lambda_{ik}$ can be used to to propagate
possible changes or updates of the input reaction rates ($R_k\to
R_k+\delta R_k$) to the theoretical abundances ($Y_i\to
Y_i+Y_i\lambda_{ik}\delta R_k/R_k$). Moreover, they enter in the
calculation of the theoretical error matrix for the abundances,
$\sigma^2_{ij}=Y_i\,Y_j\sum_k\lambda_{ik}\lambda_{jk}(\Delta\,R_k/R_k)^2$.
This matrix, summed to the experimental error matrix
$\overline\sigma^2_{ij}=\delta_{ij}\overline\sigma_i\overline\sigma_j$
and then inverted \cite{us}, defines the covariance matrix of a simple
$\chi^2$ statistical estimator. Contours of equal $\chi^2$ can then be
used to set bounds on the parameters $(x,N_\nu)$ at selected
confidence levels.

In Ref.~\cite{us} we gave polynomial fits for the functions
$Y_i(x,N_\nu)$ and $\lambda_{ik}(x,N_\nu)$ for $x\in[0,1]$ and $N_\nu=3$.  The
extension of our method to the case $N_\nu\neq3$ (say,
$1\leq\,N_\nu\,\leq5$) is, in principle, straightforward, since it
simply requires recalculation of the functions $Y_i$ and
$\lambda_{ik}$ at the chosen value of $N_\nu$. However, it would not
be practical to present, or to use, extensive tables of polynomial
coefficients for many different values of $N_\nu$. Therefore, we have
devised some formulae which, to good accuracy, relate the calculations
for arbitrary values of $N_\nu$ to the standard case $N_\nu=3$, thus
reducing the numerical task dramatically. Such approximations are
discussed in the next subsection.

\subsection{Useful Approximations}

As is known from previous work \cite{fixed}, the synthesized elemental
abundances ${\rm D}/{\rm H}$, $^3{\rm He}/{\rm H}$, and $^7{\rm
Li}/{\rm H}$ (i.e., $Y_2$, $Y_3$, and $Y_7$ in our notation) are given
to a good approximation by the quasi-fixed points of the corresponding
rate equations, which formally read
\begin{equation}
 \frac{{\rm d}Y_i}{{\rm d}t} \propto \eta\, \sum_{+,-} Y \times Y \times 
  \langle\sigma v\rangle_T\ ,
\end{equation}
where the sum runs over the relevant source $(+)$ and sink $(-)$
terms, and $\langle \sigma v\rangle_T$ is the thermally-averaged
reaction cross section. Since the temperature of the universe evolves
as $dT/dt\propto-T^3\sqrt{g_\star}$, with the number of relativistic
degrees of freedom, $g_\star=2+(7/4)(4/11)^{4/3}N_{\nu}$ (following
$e^+e^-$ annihilation), the above equation can be rewritten as
\begin{equation}
 \frac{{\rm d}Y_i}{{\rm d}T} \propto -\frac{\eta}{g_\star^{1/2}}
  \,T^{-3}\, \sum_{+,-}  Y \times Y \times \langle\sigma v\rangle_T\ ,
\end{equation}
which shows that $Y_2$, $Y_3$, and $Y_7$ depend on $\eta$ and $N_\nu$
essentially through the combination $\eta/g_\star^{1/2}$. Thus the
calculated abundances $Y_2$, $Y_3$, and $Y_7$ (as well as their
logarithmic derivatives $\lambda_{ik}$) should be approximately
constant for
\begin{equation}
 \log \eta -\frac{1}{2} \log g_\star = {\rm const}\ ,
\label{shift}
\end{equation}
as we have verified numerically.

Equation~(\ref{shift}), linearized,
 suggests that the values of $Y_i$ and of
$\lambda_{ik}$ for $N_\nu=3+\Delta{N_\nu}$ can be related to the case
$N_\nu=3$ through an appropriate shift in $x$:
\begin{eqnarray}
 Y_i(x,3+\Delta{N_\nu}) &\simeq& Y_i(x+c_i\Delta{N_\nu},3)\ ,\\
 \lambda_{ik}(x,3+\Delta{N_\nu}) &\simeq& \lambda_{ik}(x+c_i\Delta{N_\nu},3)\ ,
\end{eqnarray}
where the coefficient $c_i$ is estimated to be $\sim-0.03$ from
Eq.~(\ref{shift}) (at least for small $\Delta{N_\nu}$). In order to
obtain a satisfactory accuracy in the whole range
$(x,N_\nu)\in[0,1]\times [1,5]$, we allow upto a second-order
variation in $\Delta{N_\nu}$, and for a rescaling factor of the
$Y_i$'s:
\begin{eqnarray}
 Y_i(x,3+\Delta{N_\nu}) &=& (1+a_i\Delta{N_\nu} +b_i\Delta{N_\nu}^2)\,
  Y_i(x+c_i\Delta{N_\nu}+d_i\Delta{N_\nu}^2,3)\ , \label{sY}\\
 \lambda_{ik}(x,3+\Delta{N_\nu}) &=& \lambda_{ik}(x+
  c_i\Delta{N_\nu}+d_i\Delta{N_\nu}^2,3)\label{slambda}\ .
\end{eqnarray} 
We have checked that the above formulae (with coefficients determined
through a numerical best-fit) link the cases $N_\nu\neq 3$ to the
standard case $N_\nu=3$ with very good accuracy.

As regards the $^4{\rm He}$ abundance ($Y_4$ in our notation), a
semi-analytical approximation also suggests a relation between $x$
and $N_\nu$ similar to Eq.~(\ref{shift}), although with different
coefficients \cite{bernstein}. Indeed, functional relations of the
kind (\ref{sY},\ref{slambda}) work well also in this case. However, in
order to achieve higher accuracy and, in particular, to match the
result of the recent precision calculation of $Y_4$ which includes all
finite temperature and finite density corrections \cite{lt98}, we also
allow for a rescaling factor for the $\lambda_{4k}$'s.

We wish to emphasize that the validity of our prescription \cite{us}
for the evaluation of the BBN uncertainties and for the $\chi^2$
statistical analysis does {\em not} depend on the approximations
discussed above. The semi-empirical relations (\ref{sY},\ref{slambda})
are only used to enable us to provide the interested reader with a
simple and compact numerical code \cite{website}. 
This allows easy extraction of
joint fits to $x$ and $N_\nu$ for a given set of elemental
abundances, without having to run the full BBN code, and with no
significant loss in accuracy.

%==========================================================================
\section{Primordial Light Element Abundances}

The abundances of the light elements synthesized in the big bang have
been subsequently modified through chemical evolution of the
astrophysical environments where they are measured
\cite{chemevol}. The observational strategy then is to identify sites
which have undergone as little chemical processing as possible and
rely on empirical methods to infer the primordial abundance. For
example, measurements of deuterium (D) can now be made in quasar
absorption line systems (QAS) at high red shift; if there is a
``ceiling'' to the abundance in different QAS then it can be assumed
to be the primordial value. The helium ($^4$He) abundance is measured
in H~II regions in blue compact galaxies (BCGs) which have undergone
very little star formation; its primordial value is inferred either by
using the associated nitrogen or oxygen abundance to track the stellar
production of helium, or by simply observing the most metal-poor
objects \cite{hoganrev}. (We do not consider $^3$He which can undergo
both creation and destruction in stars \cite{chemevol} and is thus
unreliable for use as a cosmological probe.) Closer to home, the
observed uniform abundance of lithium ($^7{\rm Li}$) in the hottest
and most metal-poor Pop~II stars in our Galaxy is believed to reflect
its primordial value \cite{molarorev}.

However as observational methods have become more sophisticated, the
situation has become more, instead of less, uncertain. Large
discrepancies, of a systematic nature, have emerged between different
observers who report, e.g., relatively `high'
\cite{dhi,dhinew,dhinew2} or `low' \cite{dlo,bt98a,bt98b} values of
deuterium in different QAS, and `low' \cite{he4old,oss97} or `high'
\cite{he4new,it98} values of helium in BCG, using different data
reduction methods. It has been argued that the Pop~II lithium
abundance \cite{rbdt96,mpb95,bm97} may in fact have been significantly
depleted down from its primordial value \cite{pdd92,vc9598}, with
observers arguing to the contrary \cite{bm98}. We do not wish to take
sides in this matter and instead consider several combinations of
observational determinations, which cover a wide range of
possibilities, in order to demonstrate our method and obtain
illustrative best-fits for $\eta$ and $N_\nu$. The reader is invited
to use the programme we have provided \cite{website} to analyse other
possible combinations of observational data.

\subsection{Data Sets}

The data sets we consider are tabulated in Table~\ref{t1}. Below we
comment in detail on our choices.

\begin{itemize}

\item \underline{Data Set A:} This is taken from Ref.\cite{ot97} who
performed the first detailed MC+ML analysis to determine $\eta$ and
$N_\nu$ and is chosen essentially for comparison with our method, as
in our previous paper \cite{us}.

Their adopted value of the primordial deuterium abundance,
$\overline{Y_2}=1.9\pm0.4\times10^{-4}$, was based on early
observations of a QAS at redshift $z=3.32$ towards Q0014+813 which
suggested a relatively `high' value \cite{dhi}, and was consistent
with limits set in other QAS, but in conflict with the much lower
abundance found in a QAS at $z=3.572$ towards Q1937-1009
\cite{dlo}. More recently, observations of a QAS at $z=0.701$ towards
Q1718+4807 have also yielded a high abundance \cite{dhinew,dhinew2} as
we discuss later.

The primordial helium abundance was taken to be $\overline{Y_4}=
0.234\pm0.002\pm0.005$ from linear regression to zero metallicity in a
set of 62 BCGs \cite{oss97}, based largely on observations which gave
a relatively `low' value \cite{he4old}.

Finally the primordial lithium abundance
$\overline{Y_7}=1.6\pm0.36\times10^{-10}$ was taken from the Pop~II
observations of Ref.\cite{mpb95}, assuming no depletion.

\item \underline{Data Set B:} This corresponds to the alternative
combination of `low' deuterium and `high' helium, as considered in our
previous work \cite{us}, with some small changes.

The primordial deuterium abundance,
$\overline{Y_2}=3.4\pm0.3\times10^{-5}$, adopted here is the average
of the `low' values found in two well-observed QAS, at $z=3.572$
towards Q1937-1009 \cite{bt98a}, and at $z=2.504$ towards Q1009+2956
\cite{bt98b}.

The primordial helium abundance, $\overline{Y_4}=0.245\pm0.004$, is
taken to be the average of the values found in the two most metal-poor
BCGs, I~Zw~18 and SBS~0335-052, from a new analysis which uses the
helium lines themselves to self-consistently determine the physical
conditions in the H~II region, and specifically excludes those regions
which are believed to be affected by underlying stellar absorption
\cite{it98}. For example these authors demonstrate that there is
strong underlying stellar absorption in the NW component of I~Zw~18,
which has been included in earlier analyses \cite{he4old}.

The primordial lithium abundance
$\overline{Y_7}=1.73\pm0.21\times10^{-10}$ is from Ref.\cite{bm97},
again assuming no depletion. (Note that the uncertainty was
incorrectly reported as $\pm0.12\times10^{-10}$ in
Ref.\cite{molarorev}, as used in our previous work \cite{us}.)

\item \underline{Data Set C:} It has been suggested \cite{mesoturb}
that the discordance between the `high' and `low' values of the
deuterium abundance reported in QAS may be considerably reduced if the
analysis of the H+D profiles accounts for the correlated velocity
field of bulk motion, i.e. mesoturbulence, rather than being based on
multi-component microturbulent models. It is then found
\cite{mesoturb} that a value of
$\overline{Y_2}=(3.5-5.2)\times10^{-5}$ is compatible simultaneously
(at 95\% C.L.)
with observations of the QAS at $z=0.701$ towards Q1718+4807 (in which
a `high' value was reported \cite{dhinew,dhinew2}), and observations
of the QAS at $z=3.572$ towards Q1937-1009 and at $z=2.504$ towards
Q1009+2956 (in which a `low' value was reported \cite{bt98a,bt98b}).
We adopt this value, along with the same helium abundance as in Set B.

It has also been argued that the lithium abundance observed in Pop~II
stars has been depleted down from a primordial value of
$\overline{Y_7}=3.9\pm0.85\times10^{-10}$ \cite{pwsn98}, the lower end
of the range being set by the presence of highly overdepleted halo
stars and consistency with the $^7$Li abundance in the Sun and in open
clusters, while the upper end of the range is set by the observed
dispersion of the Pop~II abundance ``plateau'' and the $^6$Li/$^7$Li
depletion ratio. We adopt this value, noting that a somewhat smaller
depletion is suggested by other workers \cite{vc9598} who find a
primordial abundance of $\overline{Y_7}=2.3\pm0.5\times10^{-10}$.

\item \underline{Data Set D:} Recently, a `high' value of the
deuterium abundance, $\overline{Y_2}=3.3\pm1.2\times10^{-4}$, has been
reported from observations of a QAS at $z=0.701$ towards Q1718+4807
\cite{dhinew2}, in confirmation of an earlier claim \cite{dhinew}. We
adopt this value along with the same helium abundance as in set A.

For the lithium abundance, we adopt the same value \cite{bm97} as in
Set B but increase the systematic error by 0.02 dex to allow for the
uncertainty in the oscillator strengths of the lithium lines
\cite{paolo}.

\end{itemize}

\subsection{Qualitative Implications on $N_\nu$ and $\eta$}

\begin{figure}
\epsfysize=21truecm
\phantom{.}
\vspace*{0.7truecm}
\hspace*{0.1truecm}
\epsfbox{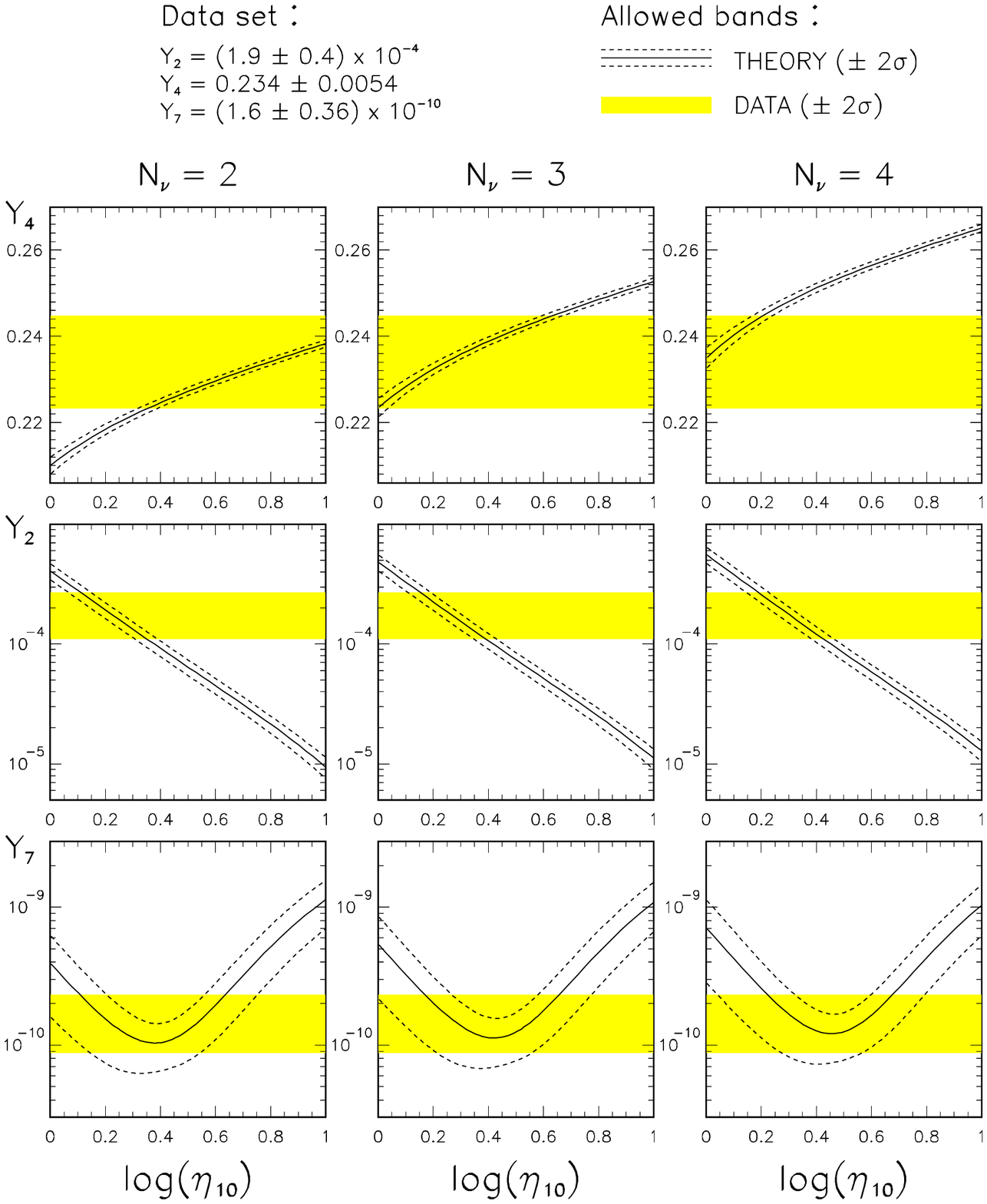}
\vskip-1.3cm
\caption{Primordial abundances $Y_4$ ($^4$He mass fraction) $Y_2$
 (D/H) and $Y_7$ ($^7$Li/H), for $N_\nu=2$, 3, and 4.  Solid and
 dashed curves represent the theoretical central values and the $\pm
 2\sigma$ bands, respectively. The grey areas represent the $\pm
 2\sigma$ experimental bands for the data set A in
 Table~\protect\ref{t1}.}
\label{fig1}
\end{figure}

\begin{figure}
\epsfysize=21truecm
\phantom{.}
\vspace*{0.7truecm}
\hspace*{0.1truecm}
\epsfbox{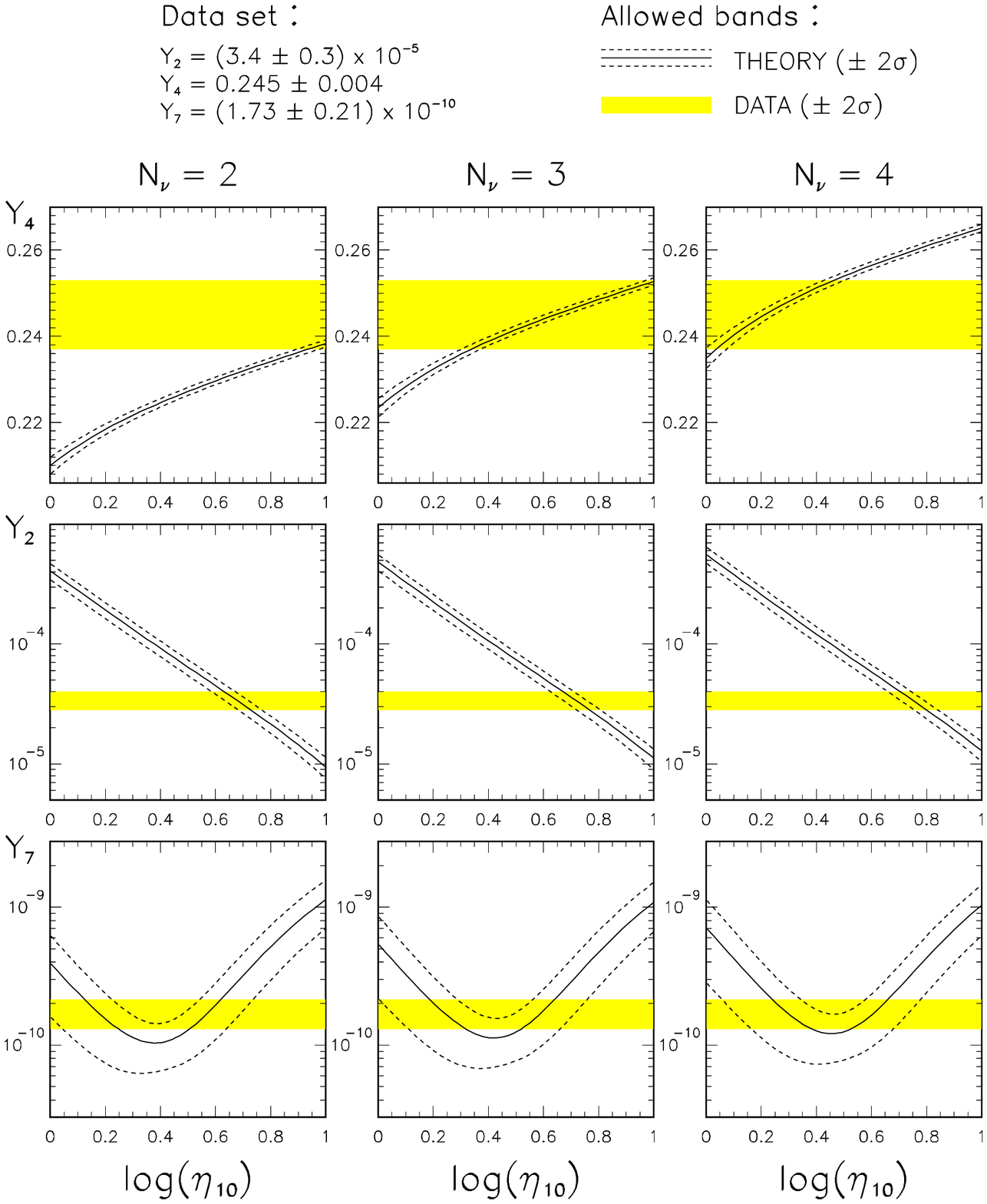}
\vskip-1.3cm
\caption{As in Fig.~\protect\ref{fig1}, but for the data set B.}
\label{fig2}
\end{figure}

\begin{figure}
\epsfysize=21truecm
\phantom{.}
\vspace*{0.7truecm}
\hspace*{0.1truecm}
\epsfbox{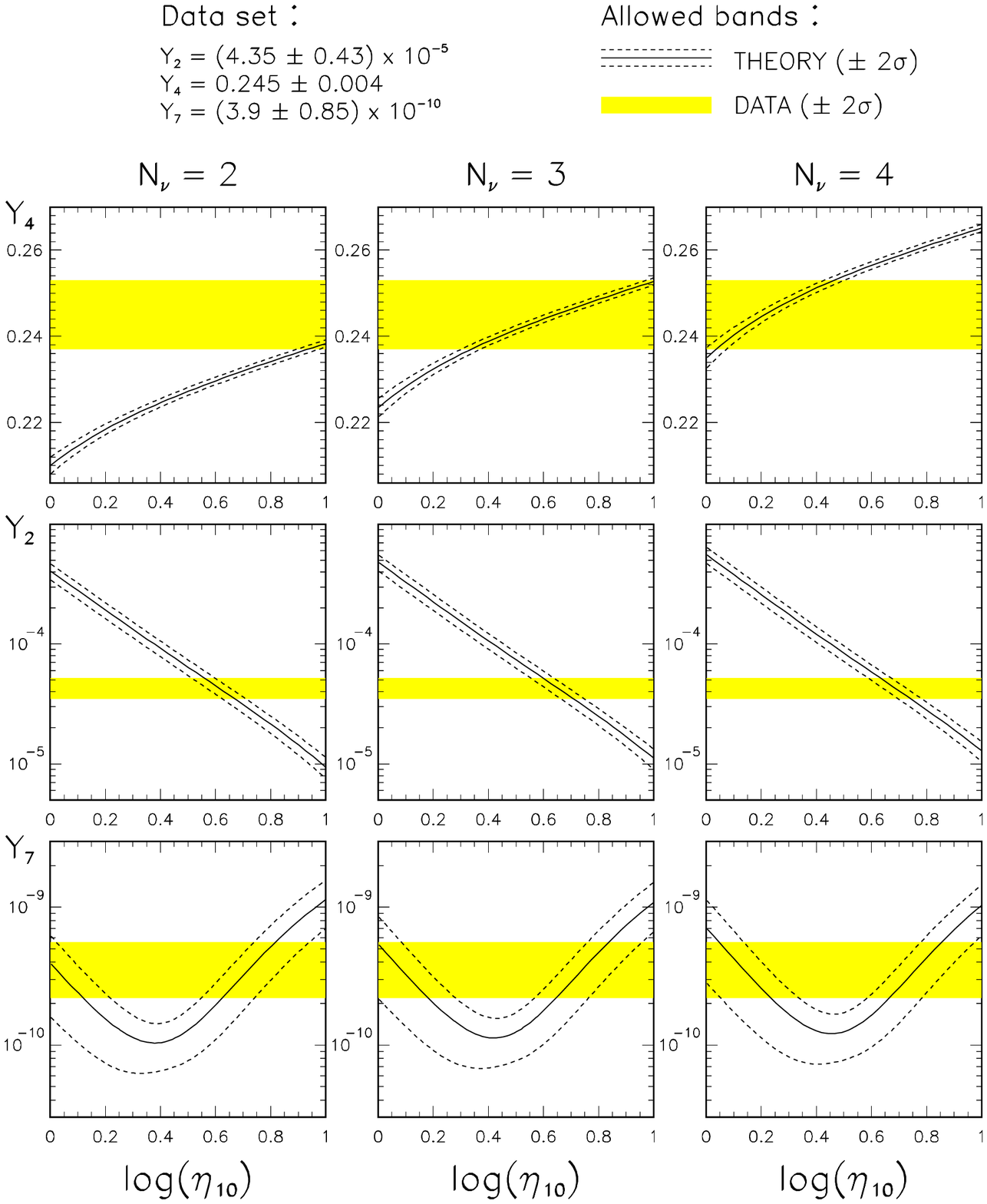}
\vskip-1.3cm
\caption{As in Fig.~\protect\ref{fig1}, but for the data set C.}
\label{fig3}
\end{figure}

\begin{figure}
\epsfysize=21truecm
\phantom{.}
\vspace*{0.7truecm}
\hspace*{0.1truecm}
\epsfbox{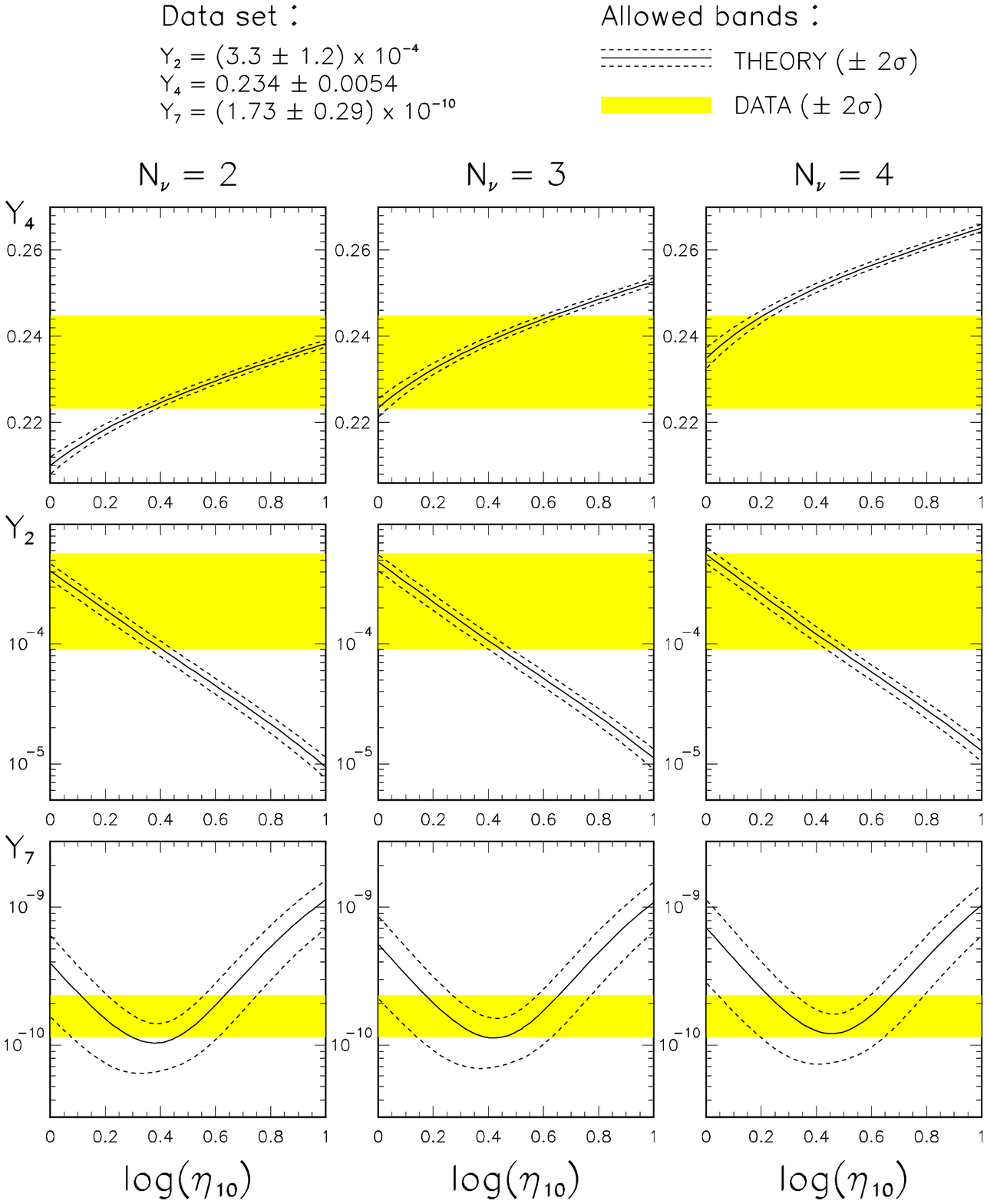}
\vskip-1.3cm
\caption{As in Fig.~\protect\ref{fig1}, but for the data set D.}
\label{fig4}
\end{figure}

Different choices for the input data sets (A--D) lead to different
implications for $\eta$ and $N_\nu$, that can be qualitatively
understood through Figs.~\ref{fig1}--\ref{fig4}.

Figure~\ref{fig1} shows the BBN primordial abundances $Y_i$ (solid
lines) and their $\pm 2\sigma$ bands (dashed lines), as functions of
$x\equiv\log(\eta/10^{-10})$, for $N_\nu=$ 2, 3, and 4. The grey areas
represent the $\pm2\sigma$ bands allowed by the data set A (see
Table~\ref{t1}). There is global consistency between theory and data
for $x\sim0.2-0.4$ and $N_\nu=3$, while for $N_\nu=2$ ($N_\nu=4$) the
$Y_2$ data prefer values of $x$ lower (higher) than the $Y_4$
data. Therefore, we expect that a global fit will favor values of
$(x,N_\nu)$ close to $(0.3,3)$.

Figure~\ref{fig2} is analogous, but for the data set B. In this case,
there is still consistency between theory and data at $N_\nu=3$,
although for values of $x$ higher than for the data set A. For
$N_\nu=2$ $(N_\nu=4)$ the combination of $Y_2$ and $Y_7$ data prefer
values of $x$ lower (higher) than $Y_4$. The best fit is thus expected
to be around $(x,N_\nu)\sim(0.7,3)$.

Similarly, Figure~\ref{fig3} shows the abundances for the data set
C. The situation is similar to data set B (Fig.~\ref{fig2}), but one
can envisage a best fit at a slightly lower value of $x$, due to the
higher value of $Y_2$, partly opposed by the increase in $Y_7$.

Finally, Fig.~\ref{fig4} refers to the data set D. In this case, data
and theory are not consistent for $N_\nu=2$, since $Y_2$ and $Y_4$
pull $x$ in different directions, and no ``compromise'' is possible
since intermediate values of $x$ are disfavored by $Y_7$. However, for
$N_\nu=3$ there is relatively good agreement between data and theory
at low $x$. Therefore, we expect a best fit around
$(x,N_\nu)\sim(0.2,3)$.

The qualitative indications discussed here are confirmed by a more
accurate analysis, whose results are reported in the next section.

%============================================================================
\section{Determining $N_\nu$}

We now present the results of fits to the data sets A--D in the
$(x,N_\nu)$ variables, using our method to estimate the correlated
theoretical uncertainties, and adopting $\chi^2$ statistics to include
both theoretical and experimental errors. We have used the theoretical
$Y_i$'s obtained by the standard (updated) BBN evolution code
\cite{code}, and checked that using the polynomial fits given in
Sec.~II~B induce negligible changes which would not be noticeable on
the plots.

In the analysis, we optionally take into account a further constraint
on $\eta$ (independent on $N_\nu$) coming from a recent analysis of
the Ly$\alpha$-``forest'' absorption lines in quasar spectra. The
observed mean opacity of the lines requires some minimum amount of
neutral hydrogen in the high redshift intergalactic medium, given a
lower bound to the flux of ionizing radiation. Taking the latter from
a conservative estimate of the integrated UV background due to
quasars, Ref.\cite{lyalpha} finds the constraint $\eta\ge
3.4\times 10^{-10}$. This bound is not well-defined statistically but,
for the sake of illustration, we have parametrized it through a
penalty function quadratic in $\eta$:
\begin{equation}
 \chi^2_{{\rm Ly}\alpha}(\eta) = 2.7 \times
  \left(\frac{3.4\times 10^{-10}}{\eta}\right)^2\ ,
\end{equation}
to be eventually added to the $\chi^2(\eta,N_\nu)$ derived from our
fit to the elemental abundances. The above function excludes values of
$\eta$ smaller than $3.4\times 10^{-10}$ at 90\% C.L. (for one degree
of freedom, $\eta$).

\begin{figure}
\epsfysize=21truecm
\phantom{.}
\vspace*{0.7truecm}
\hspace*{0.1truecm}
\epsfbox{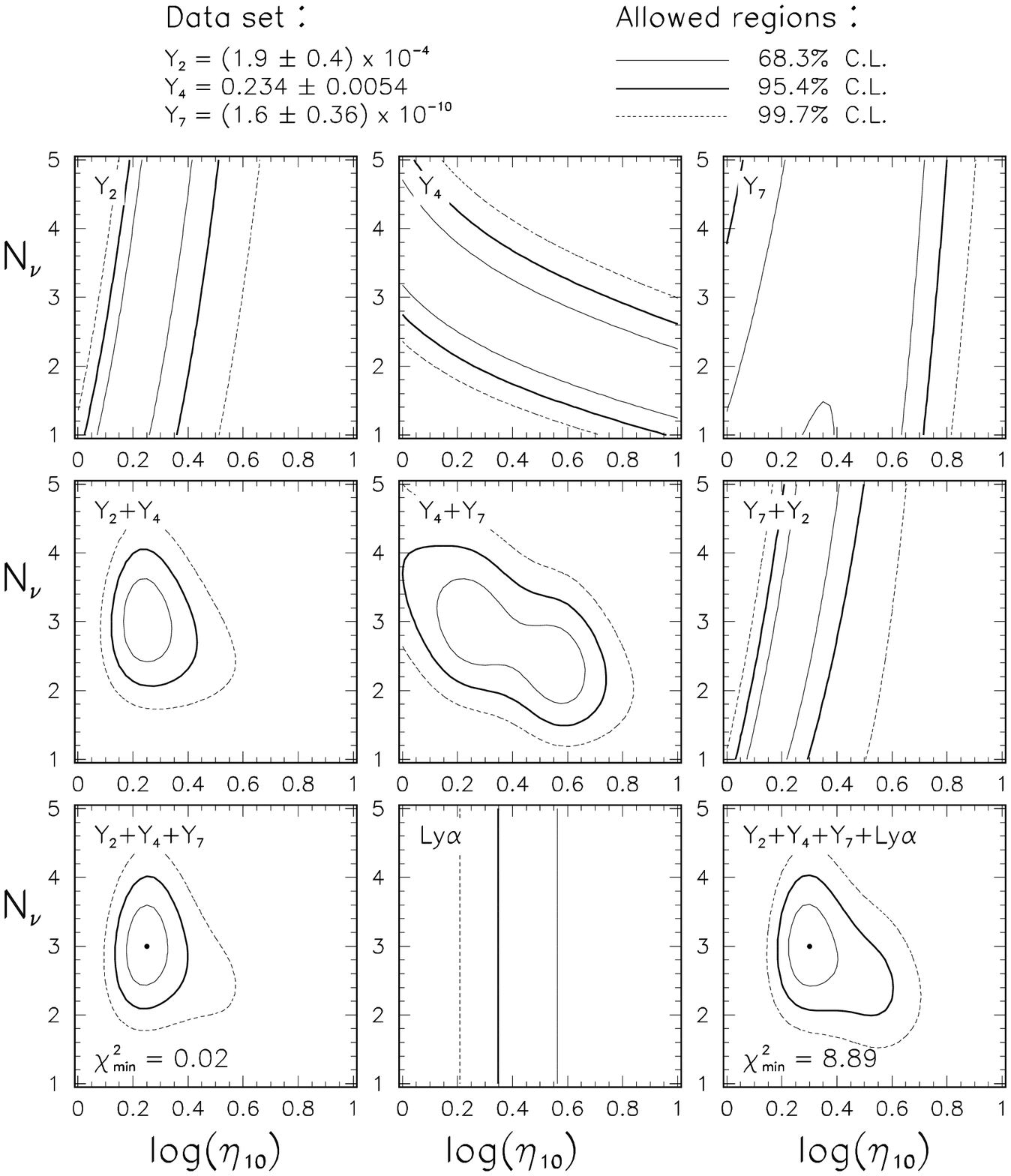}
\vskip-1.3cm
\caption{Joints fits to $x=\log(\eta_{10})$ and $N_\nu$ using the
 abundances of data set A.  The abundances $Y_2$, $Y_4$, and $Y_7$ are
 used separately (upper panels), in combinations of two (middle
 panels), and all together, without and with the Ly$\alpha$-forest
 constraint on $\eta$ (lower panels).}
\label{fig5}
\end{figure}

Figure~\ref{fig5} shows the results of joints fits to
$x=\log(\eta_{10})$ and $N_\nu$ using the abundances of data set A.
The abundances $Y_2$, $Y_4$, and $Y_7$ are used separately (upper
panels), in combinations of two (middle panels), and all together,
without and with the Ly$\alpha$-forest constraint on $\eta$ (lower
panels). In this way the relative weight of each piece of data in the
global fit can be understood at glance. The three C.L. curves (solid,
thick solid, and dashed) are defined by $\chi^2-\chi^2_{\rm min} =
2.3,\, 6.2$, and $11.8$, respectively, corresponding to 68.3\%,
95.4\%, and 99.7\% C.L. for two degrees of freedom ($\eta$ and $N_\nu$),
i.e., to the probability intervals often designated as 1, 2, and 3
standard deviation limits. The $\chi^2$ is minimized for each
combination of $Y_i$, but the actual value of $\chi^2_{\rm min}$ (and
the best-fit point) is shown only for the relevant global combination
$Y_2+Y_4+Y_7(+{\rm Ly}\alpha)$.

The results shown in Fig.~\ref{fig5} for the combinations $Y_4+Y_7$
and $Y_2+Y_4+Y_7$ are consistent with those obtained in
ref.\cite{ot97} by using the same input data but a 
completely different analysis
method (namely Monte Carlo simulation plus maximum likelihood). The
consistency is reassuring and confirms the validity of our
method. For this data set, the helium and deuterium abundances
dominate the fit, as it can be seen by comparing the combinations
$Y_2+Y_4$ and $Y_2+Y_4+Y_7$. The preferred values of $x$ are
relatively low, and the preferred values of $N_\nu$ range between 2
and 4. Although the fit is excellent, the low value of $x$ is in
conflict with the Ly$\alpha$-forest constraint on $\eta$, as indicated
by the increase of $\chi^2_{\min}$ from 0.02 to 8.89.

\begin{figure}
\epsfysize=21truecm
\phantom{.}
\vspace*{0.7truecm}
\hspace*{0.1truecm}
\epsfbox{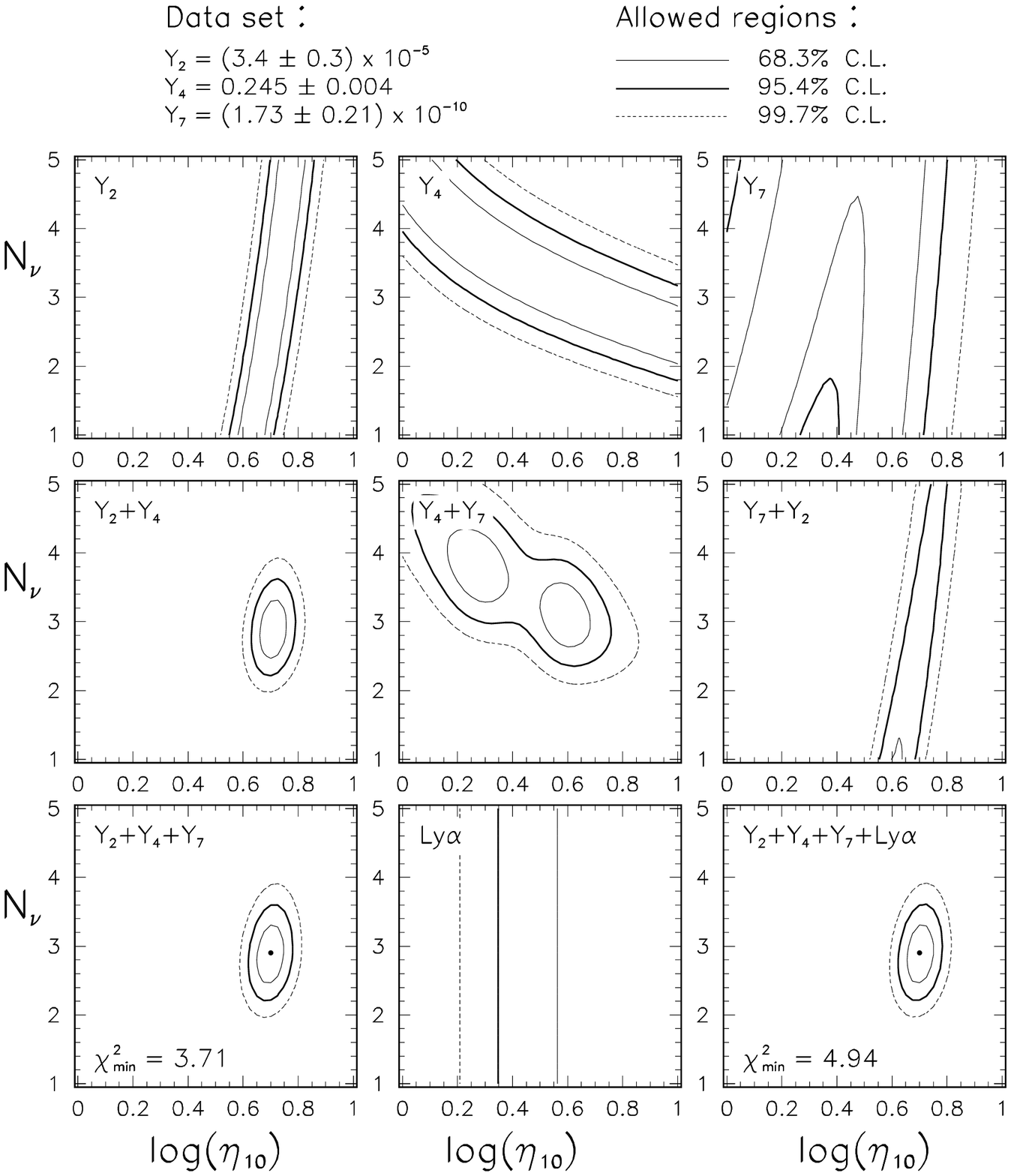}
\vskip-1.3cm
\caption{As in Fig.~\ref{fig5}, but for the data set B.}
\label{fig6}
\end{figure}

Figure~\ref{fig6} is analogous, but for the data set B which favors
high values of $x$ because of the `low' deuterium abundance. The
combination of $Y_2+Y_7$ isolates, at high $x$, a narrow strip which
depends mildly on $N_\nu$. The inclusion of $Y_4$ selects the central
part of such strip, corresponding to $N_\nu$ between 2 and 4. As in
Fig.~\ref{fig5}, the combination $Y_4+Y_7$ does not appear to be very
constraining. The overall fit to $Y_2+Y_4+Y_7$ is acceptable but not
particularly good, mainly because $Y_2$ and $Y_7$ are only marginally
compatible at high $x$. On the other hand, the $Y_2+Y_4+Y_7$ bounds
are quite consistent with the Ly$\alpha$-forest constraint.

\begin{figure}
\epsfysize=21truecm
\phantom{.}
\vspace*{0.7truecm}
\hspace*{0.1truecm}
\epsfbox{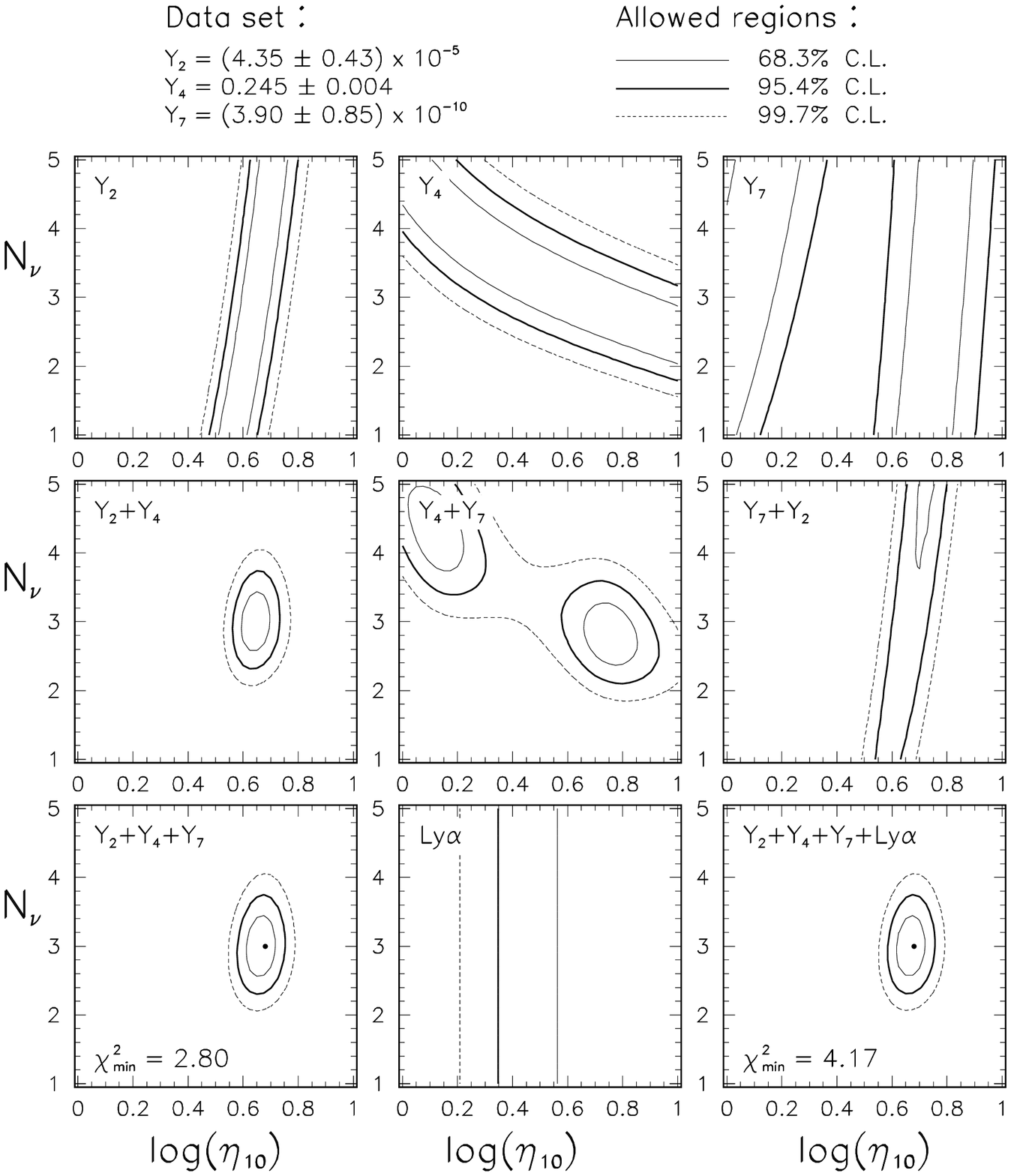}
\vskip-1.3cm
\caption{As in Fig.~\ref{fig5}, but for the data set C.}
\label{fig7}
\end{figure}

In data set C, the deuterium abundance has increased further. Moreover
the lithium abundance is no longer at the minimum of the theoretical
curve as before, so strongly disfavors ``intermediate'' values of
$x$. The overall effect, as shown in Figure~\ref{fig7}, is that
$\chi^2_{\min}$ decreases a bit with respect to Set B, and the best
fit value of $x$ moves slightly lower. The allowed value of $N_\nu$
ranges between 2 and 4. Note that had we retained the same $Y_7$ as in
Set B, then $\chi^2_{\min}$ would have dropped to 0.91 (2.55) for the
combination $Y_2+Y_4+Y_7$ (+ Ly$\alpha$-forest constraint).

\begin{figure}
\epsfysize=21truecm
\phantom{.}
\vspace*{0.7truecm}
\hspace*{0.1truecm}
\epsfbox{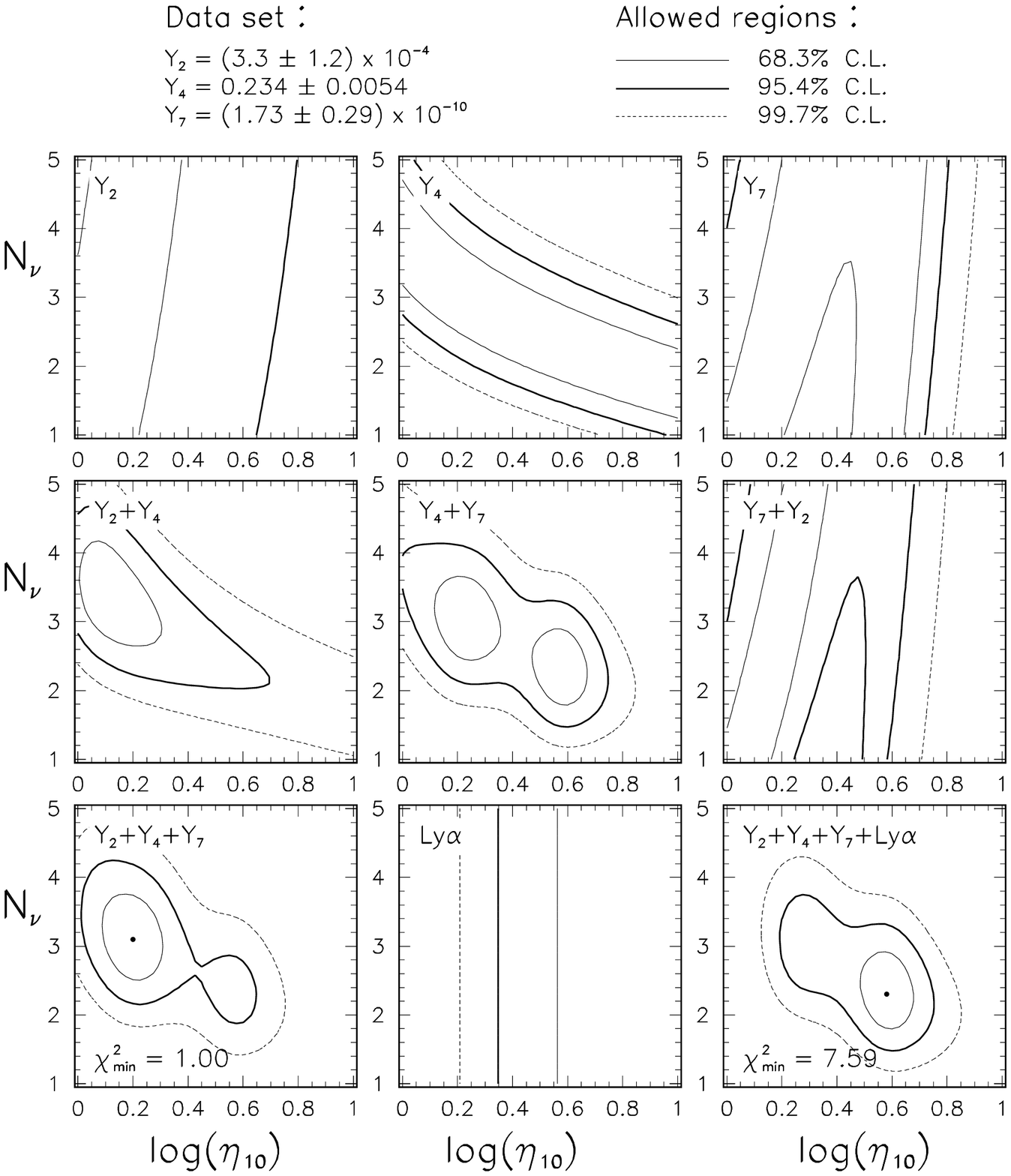}
\vskip-1.3cm
\caption{As in Fig.~\ref{fig5}, but for the data set D.}
\label{fig8}
\end{figure}

Finally, Fig.~\ref{fig8} refers to data set D which, like Data Set A,
has the `high' deuterium abundance but with larger uncertainties. So
although a low value of $x$ is still picked out, a high $x$ region is
still possible at the 2$\sigma$ level (in the $Y_2+Y_4+Y_7$ panel) and
is even favored when the Ly$\alpha$-forest bound is included (although
with an unacceptably high $\chi^2_{\min}$). Note that had the lithium
abundance been taken to be the same as in data set C (i.e. allowing
for depletion), the $\chi^2_{\min}$ would have been 0.07 (7.42) for
the combination $Y_2+Y_4+Y_7$ (+ Ly$\alpha$-forest constraint).

Of course one can also consider orthogonal combinations to those
above, e.g. `high' deuterium {\em and} `high' helium, or `low'
deuterium {\em and} `low' helium \cite{decay}. The latter combination
implies $N_\nu\sim2$, thus creating the so-called ``crisis'' for
standard nucleosynthesis \cite{crisis}. Conversely, the former
combination suggests $N_\nu\sim4$, which would also constitute
evidence for new physics. Allowing for depletion of the primordial
lithium abundance to its Pop~II value, relaxes the upper bound on
$N_\nu$ further, as noted earlier \cite{nocrisis}.

%==========================================================================
\section{Conclusions}

The results discussed above demonstrate that the present observational
data on the primordial elemental abundances are not as yet
sufficiently stable to derive firm bounds on $\eta$ and
$N_\nu$. Different and arguably equally acceptable choices for the
input data sets lead to very different predictions for $\eta$, and to
relatively loose constraints on $N_\nu$ in the range 2 to 4 at the
95\% C.L. Thus it may be premature to quote restrictive bounds based
on some particular combination of the observations, until the
discrepancies between different estimates are satisfactorily
resolved. Our method of analysis provides the reader with an
easy-to-use technique \cite{website} to recalculate the best-fit
values as the observational situation evolves further.

However one might ask what would happen if these discrepancies remain?
We have already noted the importance of an independent constraint on
$\eta$ (from the Ly$\alpha$-forest) in discriminating between
different options. However, given the many assumptions which go into
the argument \cite{lyalpha}, this constraint is rather uncertain 
at present. Fortunately
it should be possible in the near future to independently determine
$\eta$ to within $\sim5\%$ through measurements of the angular
anisotropy of the cosmic microwave background (CMB) on small angular
scales \cite{cmb}, in particular with data from the all-sky surveyors
MAP and PLANCK \cite{missions}. Such observations will also provide a
precision measure of the relativistic particle content of the
primordial plasma. Hopefully the primordial abundance of $^4$He would
have stabilized by then, thus providing, in conjunction with the above
measurements. a reliable probe of a wide variety of new physics which
can affect nucleosynthesis.

%============================================================================
\acknowledgments

We thank G. Fiorentini for useful discussions and for earlier
collaboration on the subject of this paper.

%===========================================================================

%===========================================================================
\begin{table}[h]
\caption{Experimental data sets considered in this paper for the
 elemental abundances $Y_i$.\label{t1}}
\begin{tabular}{lcccccc}
 & A&  B& C&  D\\ 
\tableline
$Y_2\times 10^5$ & $19\pm4$ & $3.4\pm0.3$ & $4.35\pm0.43$ & $33\pm12$ \\
$Y_4$  		 & $0.234\pm0.0054$ & $0.245\pm0.004$ & $0.245\pm0.004$ & 
 $0.234\pm0.0054$ \\
$Y_7\times 10^{10}$ & $1.6\pm0.36$ & $1.73\pm0.21$ & $3.9\pm0.85$ 
 & $1.73\pm0.29$
\end{tabular}
\end{table}

%========================================================================

\end{document}